\documentclass[epsfig,graphics,showpacs,twocolumn,floatfix,mathbbm,pra]{revtex4}
\usepackage{graphicx,amssymb}

\begin{document}

\newcommand{\Er}{Er$^{3+}$:Y$_{2}$SiO$_{5}$}
\newcommand{\levela}{$^{4}$I$_{15/2}$ }
\newcommand{\levelb}{$^{4}$I$_{13/2}$}
\newcommand{\system}{$\Lambda$-system}

\title{Telecommunication-wavelength solid-state memory at the single photon level}
\pacs{03.67.Hk,42.50.Gy,42.50.Md}

\author{Bj\"orn Lauritzen}
\email{bjorn.lauritzen@unige.ch}
\author{Ji\v{r}\'i Min\'{a}\v{r}}
\author{Hugues de Riedmatten}
\author{Mikael Afzelius}
\author{Nicolas Sangouard}
\author{Christoph Simon}
\author{Nicolas Gisin}

\affiliation{Group of Applied Physics, University of Geneva,
CH-1211 Geneva 4, Switzerland}

\begin{abstract} We demonstrate experimentally the
storage and retrieval of weak coherent light fields at
telecommunication wavelengths in a solid. Light pulses at the
single photon level are stored for a time up to 600 ns in an
Erbium-doped Y$_2$SiO$_5$ crystal at 2.6 K and retrieved on
demand. The memory is based on photon echoes with controlled
reversible inhomogeneous broadening, which is realized here for
the first time at the single photon level. This is implemented
with an external field gradient using the linear Stark effect.
This experiment demonstrates the feasibility of a solid state
quantum memory for single photons at telecommunication
wavelengths, which would represent an important resource in
quantum information science.
\end{abstract}
\date{\today}
 \maketitle

Quantum memories allowing the reversible transfer of quantum
states between light and matter are an essential requirement in
quantum information science \cite{Hammerer2008}. They are, for
example, a crucial resource for the implementation of quantum
repeaters\cite{Briegel1998,Duan2001,Simon2007,Sangouard2009a},
which are a potential solution to overcome the limited distance of
quantum communication schemes due to losses in optical fibers.
Several schemes have been proposed to implement photonic quantum
memories
\cite{Fleischhauer2000,Kozhekin2000,Moiseev2001,Kraus2006,Nunn2007,Sangouard2007,Afzelius2009a}.
Important progress has been made during the last few years,
with proof of principle demonstrations in atomic gases
\cite{Chaneliere2005,Eisaman2005,Choi2008,Akiba2009,Julsgaard2004,Appel2008,Honda2008,Cviklinski2008},
single atoms in a cavity \cite{Boozer2007}, and solid state
systems \cite{Riedmatten2008}. For all these experiments the
wavelength of the stored light was close to the visible range and
thus not suited for direct use in optical telecom fibers. The
ability to store and retrieve photons at telecommunication
wavelengths (around 1550 nm) in a quantum memory would provide an
important resource for long distance quantum communication. Such a
quantum memory could easily be integrated in the fiber
communication network. In combination with a photon pair source,
it could provide a narrow band triggered single photon source
adapted to long distance transmission. Moreover, quantum memories
at telecommunication wavelengths are required for certain efficient
quantum repeater architectures
\cite{Sangouard2007a,Sangouard2008,Sangouard2009a}.

A telecom quantum memory requires an atomic medium with an optical
transition in the telecom range, involving a long lived atomic
state. The only candidate proposed so far is based on erbium doped
solids, which have a transition around 1530nm between the ground
state \levela and the excited state \levelb. These systems have
been studied for spectroscopic properties
\cite{Bottger2006,Bottger2009} and classical light storage
\cite{Baldit2005,Staudt2007,Staudt2007a}. Photonic quantum storage
in these materials is extremely challenging, because of the
difficulties in the memory preparation using optical pumping
techniques \cite{Lauritzen2008}. Yet in this paper, we report an
experiment of storage and retrieval of weak light fields at the
single photon level in an erbium doped solid.

Rare-earth doped solids have an inhomogeneously broadened
absorption line. Single photons can be mapped onto this optical
transition, leading to single collective optical excitations
\cite{Riedmatten2008}. During the storage, inhomogeneous dephasing
takes place, preventing an efficient collective re-emission of the
photon. This dephasing can be compensated for using photon echo
techniques. The storage of quantum light (e.g. single photons) is
however not possible using traditional photon echo techniques,
such as two pulse photon echoes \cite{Kurnit1964}. The main issue
is that the application of the strong optical pulse ($\pi$-pulse)
to induce the rephasing mechanism leads to amplified spontaneous
emission and reduce the fidelity of the storage to an unacceptable
level \cite{Ruggiero2009}. A way to overcome this problem is to
induce the rephasing of the atomic dipoles by generating and
reversing an artificial inhomogeneous broadening. This scheme is
known as Controlled Reversible Inhomogeneous Broadening (CRIB)
\cite{Moiseev2001,Kraus2006,Alexander2006,Tittel2009}. In
rare-earth doped solids with a permanent dipole moment, this can
be done with an electric field gradient using the linear Stark
effect. The CRIB scheme was first demonstrated with bright optical
pulses , in a Eu$^{3+}$:Y$_{2}$SiO$_{5}$ crystal at 580 nm
\cite{Alexander2006}. The phase of the stored light pulses was
shown to be well preserved \cite{Alexander2007a}. For these
experiments, the storage and retrieval efficiency was of the order
of $10^{-6}$. It has been dramatically improved in more recent
experiments at 606 nm in Pr$^{3+}$:Y$_{2}$SiO$_{5}$
\cite{Hetet2008}. CRIB has also been demonstrated on a spin
transition in a rubidium vapor \cite{Hetet2008a} at 780 nm. Here,
we report an experiment at telecommunication wavelength. Moreover,
we also report the first experimental demonstration of CRIB at the
single photon level, opening the road to the quantum regime.

\begin{figure}[h]
\includegraphics[width=.45\textwidth]{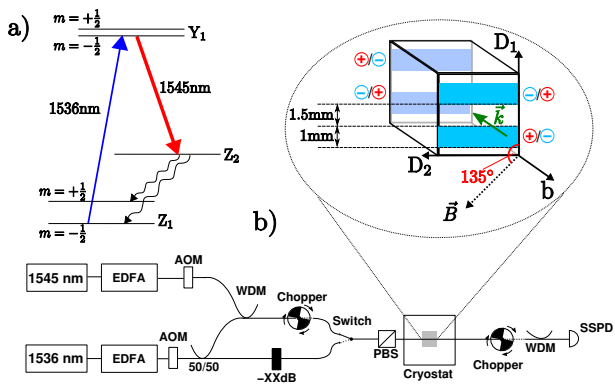}
\caption{(color online) (a) Level scheme of \Er. (b) Experimental
setup: The pump laser (external cavity diode laser at 1536nm) is
split into two paths, one for the preparation pulses and one for
the weak pulses to be stored. Pulses are created with
acousto-optical modulators (AOMs). In the preparation path, the
stimulation laser (DFB laser diode at 1545nm+fiber amplifier) is
added using a wavelength division multiplexer (WDM). The pulses to
be stored are attenuated to the single photon level with a fiber
attenuator. An optical switch allows us to send either of them
into the sample. In order to protect the detector (SSPD) and to
avoid noise from a leakage of the optical switch, two mechanical
choppers are used. The polarization of the light is aligned to
maximize the absorption using the polarizing beam splitter (PBS).
Inset: illustration of the crystal with electrodes, magnetic field
and light propagation directions indicated.}\label{setup}
\end{figure}

In order to realize a CRIB experiment in a rare-earth doped solid,
one first has to prepare a narrow absorption line within a large
transparency window. The spectrum of this line is then broadened
by an electric field gradient using the linear Stark effect to
match the bandwidth of the photon to be stored. The incident
photon is absorbed by the ions in the broadened line, and mapped
into a single collective atomic excitation. During a time $t$ each
excited ion $i$ will acquire a phase $\Delta_{i}t$ due to its
shift in the absorption frequency $\omega_{i}=\omega_{0} +
\Delta_{i}$ from the central frequency $\omega_{0}$. Switching the
polarity of the field after a time $t=\tau$ will reverse the
broadening ($\omega_{i}=\omega_{0} - \Delta_{i}$) and after
another time $\tau$ the ions will be in phase again and re-emit
the photon. In order to create the initial narrow absorption line,
a population transfer between two ground states (in our case
Zeeman states) using optical pumping via the excited state is used
\cite{Lauritzen2008}. In case of imperfect optical pumping there
will be population remaining in the excited state after the
preparation sequence. An experimental issue arising when input
pulses are at the single photon level is the fluorescence from the
remaining excited atoms. If the depletion of this level is slow
(as in rare-earth ions, with optical relaxation times $T_{1}$
usually in the range of 0.1 ms to 10 ms), this can lead to a high
noise level that may blur the weak echo pulse. The problem is
especially important for erbium doped solids, where $T_1$ is very
long ($\approx$
 11 ms \cite{Bottger2006} in \Er). In our experiment, the population
transfer is enhanced by stimulating ions from the excited state
down to the short lived second ground state crystal field level
using a second laser at 1545 nm (Fig. \ref{setup}a)
\cite{Lauritzen2008}. The application of this laser enhances the
rate of depletion of the excited state and thus reduces the noise
from fluorescence. Together with a suitable waiting time between
the preparation and the light storage, it allows the realization
of the scheme at the single photon level.

Our memory consists of an Y$_{2}$SiO$_{5}$ crystal doped with
erbium (10ppm) cooled to 2.6 K in a pulse tube cooler (Oxford
Instruments). The crystal has three mutually perpendicular
optical-extinction axes labelled D1, D2, and b. Its dimensions are
$ 3.5 \times 4 \times 6$mm along these axes. The magnetic field of
$B=$1.5 mT used to induce the Zeeman splitting necessary for the
memory preparation is provided by a permanent magnet outside the
cryostat and is applied in the $D_{1}-D_{2}$ plane at an angle of
$\theta=135^{\circ}$ with respect to the $D_{1}$-axis
\cite{Hastings-Simon2008a} (Fig.\ref{setup}b). The light is
travelling along b. The electrical field gradient for the Stark
broadening is applied with four electrodes placed on the crystal
in quadrupole configuration, as shown in Fig.\ref{setup}b and
described in \cite{Minar2009a}. The induced broadening is
proportional to the voltage $U$ applied on the electrodes
\cite{Minar2009a}.

The experiment is divided into two parts: the preparation of the
memory and the storage of the weak pulses (see Fig.\ref{setup}b).
Each preparation sequence takes 120ms of optical pumping during
which both the pump and the stimulation lasers are sent into the
sample. The frequency of the pump laser is repeatedly swept to
create a large transparency window into the inhomogeneously
broadened absorption line. If the laser is blocked for a short
time at the center of each sweep using an acousto-optical
modulator (AOM), a narrow absorption feature is left at the center
of the pit \cite{Lauritzen2008}. The time available to perform the
memory protocol is given by the Zeeman lifetime of $T_{Z}=130$ms
\cite{Hastings-Simon2008a} of the material. In order to deplete
the excited state the laser at 1545nm is left on for 23.5ms after
the pump pulses. Then the preparation path is closed and the
detection path is opened using an optical switch and mechanical
choppers. The storage sequence begins 86ms after the pump pulse,
in order to avoid fluorescence from the excited atoms. It is
composed of 8000 independent trials separated by 5$\mu s$. In each
trial, a weak pulse of duration $\delta \tau$ is stored and
retrieved. The initial peak is broadened with an electrical pulse
before the absorption of each pulse. The polarity of the field is
then inverted at a programable time after the storage, allowing
for on demand read-out. The whole sequence is repeated at a rate
of 3Hz. The weak output mode is detected using a superconducting
single photon detector (SSPD) \cite{Gol'tsman2001} with an
efficiency of 7\% and a low dark count rate of $10\pm5$Hz. The
incident pulses are weak coherent states of light
$|\alpha\rangle_{L}$ with a mean number of photons
$\overline{n}=|\alpha|^{2}$. We determined $\overline{n}$ at the
input of the cryostat by measuring the number of photons arriving
at the SSPD (with the laser out of resonance), compensating for
the transmission(16$\%$) and detection efficiency.

\begin{figure}[h]
\includegraphics[width=.38\textwidth]{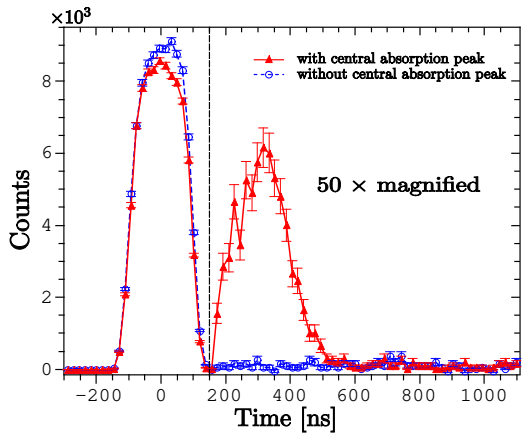}
\caption{(color online) CRIB measurement with (red triangles,
solid line) and without (blue circles, dashed line) absorption
peak, with $\overline{n}=10$ and $\delta \tau =$ 200 ns. The pulse
on the left is the transmitted part of the incident photons. One
can clearly see that the absorption is enhanced in the presence of
a peak. The electric field ($U$=$\pm$50 V) was reversed just after
the input pulse. Dark counts have been subtracted from the data.
Integration time for both curves was 200s.}\label{peak_no_peak}
\end{figure}

\begin{figure}[h]
\includegraphics[width=240pt]{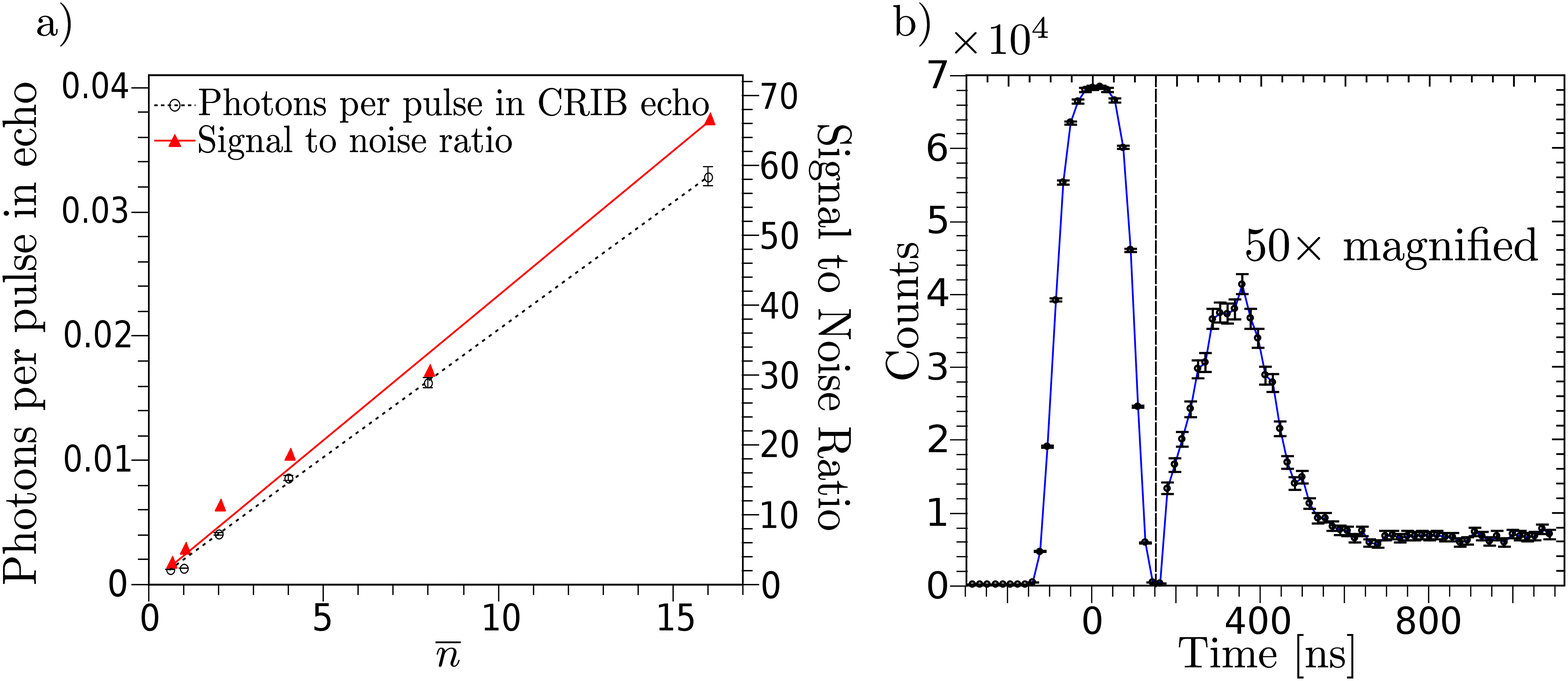}
\caption{(color online) (a) Number of photons in the CRIB echo
(open circles) and signal to noise ratio (plain triangles) as a
function of the number of incident photons $\overline{n}$, for 200
ns input pulses. (b) CRIB echo for $\overline{n}=0.6$ (integration
time 25000s). Dark counts have been subtracted from the
data.}\label{linearity}
\end{figure}

We now describe the observation of CRIB photon echoes of weak
pulses. As a first experiment, we sent pulses with $\overline{n}$
=10 and $\delta \tau$ = 200 ns into the sample. The polarity of
the electric field ($U=\pm$ 50 V) was reversed directly after each
pulse. Fig. \ref{peak_no_peak} shows a time histogram of the
photon counts detected after the crystal. The first peak
corresponds to the input photons transmitted through the crystal.
The second peak is the CRIB echo. It is clearly visible above the
noise floor. Only a small fraction of the incident light is
re-emitted in the CRIB echo (about 0.25 $\%$). The reasons for
this low storage and retrieval efficiency and ways to improve it
will be discussed in more detail below. As a consistency check, we
verified that the echo disappears when the narrow absorption peak
is not present (see blue open circles in Fig. \ref{peak_no_peak}).
\begin{figure}[h]
\includegraphics[width=210pt]{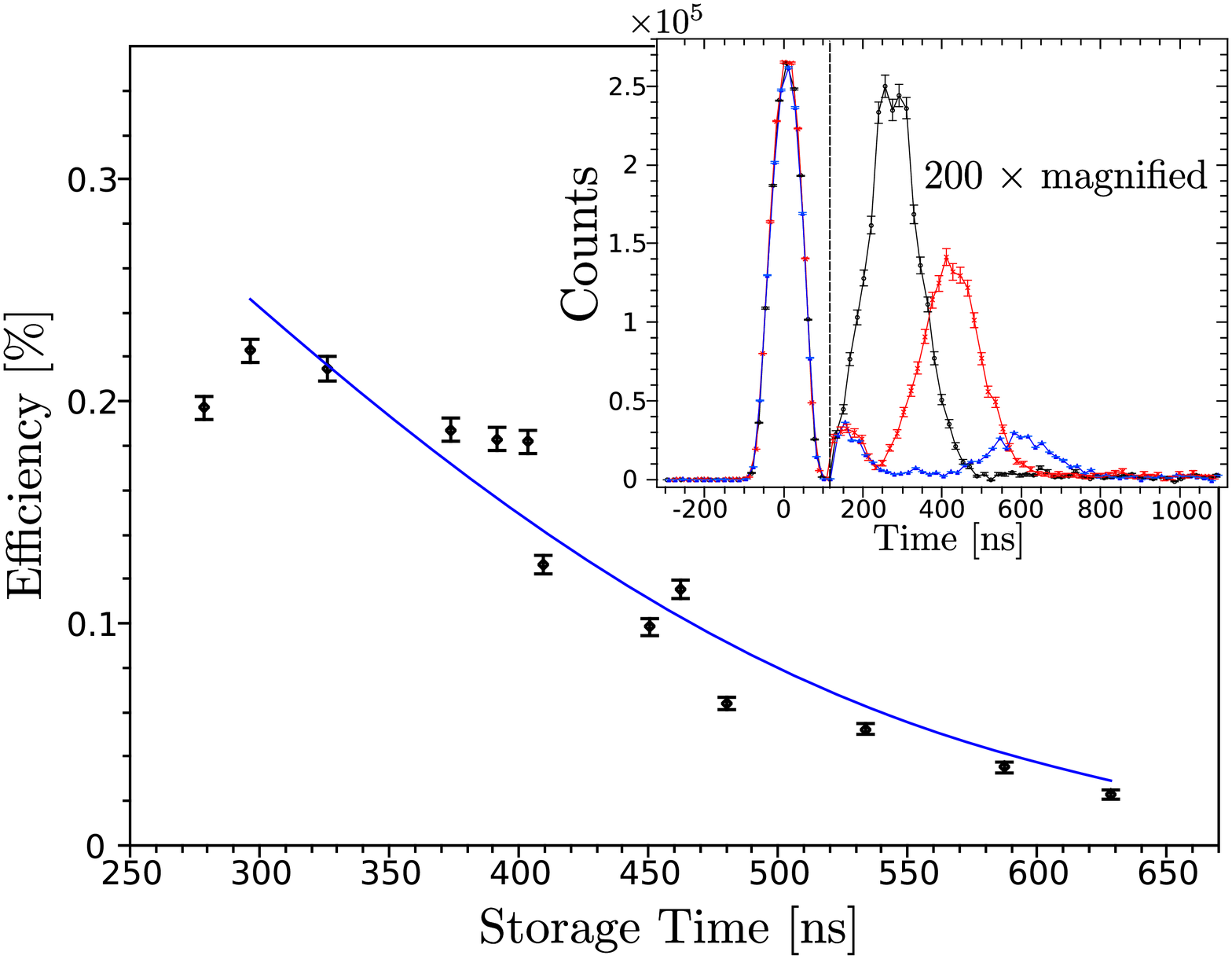}
\caption{(color online) Efficiency of the CRIB memory as a
function of storage time, for input pulses with $\delta \tau$=200
ns , $\overline{n}=10$ and $U=\pm$50 V. The error bars correspond
to the statistical uncertainty of the measured photon numbers. The
solid line is a gaussian fit (with the first point excluded). The
inset shows CRIB echoes for three different switching times of the
electrical field. For the inset curves, $\delta \tau$=100 ns,
$U$=$\pm$70 V and the integration time is 500 s.}\label{decay}
\end{figure}
By reversing the electric field gradient at later times, it is
possible to choose the retrieval time of the stored light.
Fig. \ref{decay} shows the efficiency of the CRIB echo for
different storage times. The signal was clearly visible up to
 a storage time of around 600 ns. The decay of the efficiency is
due to the finite width of the initial (unbroadened) peak
\cite{Sangouard2007} (see below). The solid line is a fit assuming
a gaussian shape for the absorption line, giving a decay time of
370 ns. Shorter pulses with $\delta \tau$ = 100 ns have also been
stored (see the inset of Fig. \ref{decay}) with a larger
broadening ($U=\pm$ 70 V), leading to a larger time-bandwidth
ratio, with however a reduced storage efficiency. Finally, we
gradually lowered $\overline{n}$ by increasing the attenuation,
for input pulses with $\delta \tau$=200 ns . The result is shown
in Fig. \ref{linearity}a. Both the number of photons in the CRIB
echo and the signal to noise ratio depend linearly on
$\overline{n}$. This means that the efficiency and the noise are
independent of $\overline{n}$.
 Fig. \ref{linearity}b shows the result of a measurement with - in
quantum key distribution terminology - pseudo-single photons
($\overline{n}=0.6$ photons per pulse). In that case, we still
obtain a signal to noise ratio of $\sim$3. The remaining noise
floor may be due to residual fluorescence and leakage through the
AOM creating the pulses to be stored.

In the following we analyze the efficiency and storage time
performances of our memory in more detail. Ref.
\cite{Sangouard2007} gives a simplified model for the CRIB memory.
In this model, the storage and retrieval efficiency if the echo is
emitted in the forward direction is given by :
\begin{equation}\label{decay_equ}
\eta_{CRIB}(t)=d^{2}e^{-d}e^{-t^{2}\tilde{\gamma}^{2}} \label{eff}
\end{equation}
where $\tilde{\gamma}=2\pi\gamma$ is the spectral width (standard
deviation) of the initial gaussian absorption peak, and $d$ is the
optical depth of the broadened absorption peak. The main
assumption here is that the spectral width of the absorption peak
is much wider than the spectral bandwidth of the photon to be
stored.
By
fitting the decay curve of Fig.\ref{decay} with
Eq.\ref{decay_equ},  we find a full width at half maximum
linewidth of the central peak of 1MHz. This corresponds well to
the results obtained by a measurement of the transmission
spectrum.
 The minimal width is limited
by the linewidth of our unstabilized  laser diode and power
broadening during the preparation of the peak. The optical
coherence time of the transition under our experimental conditions
has been measured independently by photon echo spectroscopy. It
was found to be $T_{2}\approx2\mu s$, corresponding to a
homogeneous linewidth of 160 kHz. Note that the optical coherence
in \Er  could be drastically increased using lower temperatures
and higher magnetic fields \cite{Bottger2009}.

In our experiment, imperfect optical pumping results in a large
absorbing background with optical depth $d_0$, which acts as a
passive loss, such that the experimental storage and retrieval
efficiency is given by: $\eta(t)=\eta_{CRIB}(t)$exp$(-d_0)$
\cite{Riedmatten2008}. The values of $d$ and $d_0$ can be measured
by recording the absorption spectra. This yields an optical depth
of the unbroadened peak $d'=0.5 \pm 0.2$ and an absorbing
background of $d_{0}=1.6 \pm 0.1$. A voltage of 50 V on the
electrodes corresponds to a broadening of a factor of $\sim$3
\cite{Minar2009a}, leading to $d=0.17\pm 0.07$. In our experiment,
the photon bandwidth is of the same order as the broadened peak,
so that the assumptions of the simplified model are not fulfilled.
In order to have a more accurate description, we have solved
numerically the Maxwell-Bloch equations with the measured $d'$,
using a gaussian initial peak. This gives storage and retrieval
efficiencies of order $1.5\cdot 10^{-3}$ (including the passive
loss $d_0$) for a storage time of 300 ns, in reasonable agreement
with the measured values (see Fig. \ref{decay})

This study shows that the main reason of the low storage and
retrieval efficiency in the present experiment is the small
absorption in the broadened peak and the large absorbing
background, due to imperfect optical pumping. About 80 $\%$ of the
retrieved photons are lost in the absorbing background. The
limited optical pumping efficiency is due to the small branching
ratio in the $\Lambda$ system and to the small ratio between the
relaxation life times of the optical and the ground state Zeeman
transitions \cite{Lauritzen2008}. This could be improved in
several ways. First technical improvements can be implemented,
such as using lower temperatures, higher stimulation laser
intensities and spin mixing in the excited state using RF fields,
as demonstrated in \cite{Lauritzen2008}. Second, the branching
ratio and Zeeman life time strongly depend on the applied magnetic
field angle and intensity. A full characterization of the optical
pumping efficiency with respect to these parameters has not been
carried out yet. Finding optimal conditions may lead to
significant improvements. It would also be interesting to
investigate hyperfine states. Finally, other crystals might be
explored, e.g., Y$_{2}$O$_{3}$, to search for longer Zeeman
lifetimes.

In summary, we have presented a proof-of-principle of quantum
memory for photons at telecommunication wavelengths. Pulses of
light at the single photon level have been stored and retrieved in
an Erbium doped crystal, using the CRIB protocol. Continuing
efforts to increase the efficiency and the storage time will be
required in order to build a useful device for applications in
quantum information science. Our experiment is nevertheless an
enabling step towards the demonstration of a fiber network
compatible quantum light matter interface. It also confirms the
feasibility of the CRIB protocol at the single photon level.

The authors acknowledge technical assistance by Claudio Barreiro
and Jean-Daniel Gautier as well as stimulating discussions with
Imam Usmani, Wolfgang Tittel, Sara Hastings-Simon, Matthias Staudt
and Nino Walenta. This work was supported by the Swiss NCCR
Quantum Photonics, by the European Commission under the Integrated
Project Qubit Applications (QAP) and the ERC-AG Qore.


\end{document}